\documentclass{IEEEtran}
\usepackage{array}
\usepackage{cite}
\usepackage{amsmath,amssymb,amsfonts}
\usepackage{algorithmic}
\usepackage{geometry}
\usepackage{graphicx}
\usepackage{float}
\usepackage{caption}
\usepackage{subfigure}
\usepackage{textcomp}
\usepackage{xcolor}
\usepackage{nicematrix}
\usepackage{booktabs}
\usepackage{cuted} \stripsep -3pt plus 3pt minus 2pt
\def\BibTeX{{\rm B\kern-.05em{\sc i\kern-.025em b}\kern-.08em
    T\kern-.1667em\lower.7ex\hbox{E}\kern-.125emX}}
\begin{document}

\title{A Wideband MIMO Channel Model for Aerial Intelligent Reflecting Surface-Assisted Wireless Communications\\ }

\author{\IEEEauthorblockN{Shaoyi Liu\textsuperscript{1}, Nan Ma\textsuperscript{1}, Yaning Chen\textsuperscript{1}, Ke Peng\textsuperscript{1} and Dongsheng Xue\textsuperscript{2}}\\
	\IEEEauthorblockA{\textsuperscript{1}State Key Laboratory of Networking and Switching Technology, \\Beijing University of Posts and Telecommunications, Beijing, China\\
		\textsuperscript{2}Key Laboratory of Universal Wireless Communications, Ministry of Education\\Beijing University of Posts and Telecommunications, Beijing, China \\
		Email:\{sy\_liu,manan\}@bupt.edu.cn}}
\maketitle
\pagestyle{empty}  
\thispagestyle{empty} 
\begin{abstract}
Compared to traditional intelligent reflecting surfaces (IRS), aerial IRS (AIRS) has unique advantages, such as more flexible deployment and wider service coverage. However, modeling AIRS in the channel presents new challenges due to their mobility.  In this paper, a three-dimensional (3D) wideband channel model for AIRS and IRS joint-assisted multiple-input multiple-output (MIMO) communication system is proposed, where considering the rotational degrees of freedom in three directions and the motion angles of AIRS in space. Based on the proposed model, the channel impulse response (CIR), correlation function, and channel capacity are derived, and several feasible joint phase shifts schemes for AIRS and IRS units are proposed. Simulation results show that the proposed model can capture the channel characteristics accurately, and the proposed phase shifts methods can effectively improve the channel statistical characteristics and increase the system capacity. Additionally, we observe that in certain scenarios, the paths involving the IRS and the line-of-sight (LoS) paths exhibit similar characteristics. These findings provide valuable insights for the future development of intelligent communication systems.
\end{abstract}

\begin{IEEEkeywords}
	channel model, aerial intelligent reflecting surface (AIRS), IRS, propagation characteristics
\end{IEEEkeywords}

\section{Introduction}
One of the visions of the sixth generation (6G) wireless communication is to enhance the Internet of Things (IoT) by expanding its coverage and capabilities, achieving intelligent interconnection between different devices. Intelligent reflecting surfaces (IRS) are considered one of the key technologies for the 6G wireless communications as they have the ability to alter the propagation environment and expand network coverage \cite{jian2022reconfigurable}. IRS is composed of sub-wavelength units with tunable amplitude, phase, frequency, and other characteristics, whose size is typically between 1/10 to 1/2 of the wavelength\cite{tang2020wireless}. To better design IRS-assisted communication systems, establishing a wireless channel model is an effective method.

The research on the path loss model of the IRS channel has been accumulated. In \cite{tang2020wireless} and \cite{tang2022path}, they investigate the relationship between free space path loss and the size of the IRS unit, as well as the near-field/far-field effects of the IRS and the radiation patterns of antennas used in IRS-assisted communication. However, it is important to note that these studies utilize deterministic channel modeling to describe the performance of IRS in large-scale fading. This approach requires a significant amount of measurements. In contrast, the geometry-based stochastic model (GBSM) is widely employed to describe propagation environments due to its accuracy and versatility. In \cite{xiong2021statistical}, researchers proposed a general wideband non-stationary channel model for IRS-assisted MIMO communication scenarios. Additionally, in \cite{jiang2021general}, they presented a 3D non-stationary channel model based on ellipsoids for IRS-assisted MIMO wideband communication. Furthermore, in \cite{xiong20223d}, an unmanned aerial vehicle (UAV)-to-ground model for IRS-assisted communication was proposed.

While previous works have extensively investigated the impact of IRS on terrestrial wireless channels, there is a growing interest in exploring the application of IRS in the context of space-air-ground integrated networks (SAGIN), which are considered a key component of 6G networks \cite{zhang20196g}. UAVs play a pivotal role in SAGIN and can be used to deploy IRS on high-altitude platforms (HAPs) or UAVs themselves, creating what is referred to called aerial IRS (AIRS). Compared with terrestrial IRS, AIRS offers several advantages such as a broader service range, easier establishment of line-of-sight (LoS) links with users, and more flexibility in adjusting angles and positions. These attributes make AIRS more suitable for serving mobile users \cite {lu2021aerial}. However, it is worth noticing that the research on AIRS models is still in its early stages. One of the initial studies by Ma et al.\cite {ma2021multipath} proposed a narrowband model for AIRS-assisted communication system, discussing the channel impulse response (CIR) and Doppler shifts, subsequent research conducted in \cite {ma2022modeling} focused on exploring the statistical characteristics of channel and investigating phase shifts methods. However, these studies did not consider the complex motion of UAVs, and the use of narrowband models may not be sufficient for addressing the requirements of future 6G communication systems.

To address the aforementioned gap, this paper proposes a novel 3D wideband channel model for AIRS and IRS joint-assisted MIMO communication, which takes into account the impact of terrestrial IRS cooperative assistance and the AIRS movement on propagation characteristics. This model considers the rotational degrees of freedom in three directions and the motion angles of AIRS in space. Additionally, the model employs a cluster delay line (CDL) structure to accurately simulate wideband channel characteristics. From this model, the CIR, correlation function, and channel capacity are derived. Furthermore, a phase shifts design method for the cooperative use of AIRS and IRS is proposed, which considers the impact of non-ideal IRS on the channel. The simulation results show that the proposed method can improve the channel capacity of the communication system.

The remaining sections of the paper are structured as follows: Section II describes the proposed model and the phase shifts methods of IRS. Section III studies the statistical characteristics of the proposed model. The numerical results and analysis are presented in Section IV, and finally, the paper concludes with section V.
\section{3D Wideband Channel Model for AIRS-Assisted MIMO Communication}

\subsection{Description of the Channel Model}

Fig. 1 illustrates the geometric structure of the proposed model, with the transmitter (Tx) and receiver (Rx) representing the base station (BS) and the mobile station (MS), respectively. The Tx is equipped with $P$ transmit omnidirectional antennas, where the adjacent antenna elements are separated by ${{\delta }_{T}}$. Similarly, the Rx is equipped with $Q$ receive omnidirectional antennas, where the adjacent antenna elements are separated by ${{\delta }_{R}}$. As shown in Fig. 2 (a), the azimuth and elevation orientation angles of the Tx antennas are denoted by ${{\alpha }^{T}}$ and ${{\beta }^{T}}$, respectively. The distance vector from the first antenna element to the $p$-th ($p=1,2,\ldots ,P$) Tx antenna denoted by $\pmb{An}_{p}^{T}$, can be expressed as
\begin{equation}
	\pmb{An}_{p}^{T}=(p-1){{\delta }_{T}}\left[ \begin{matrix}
		-\cos {{\beta }^{T}}\cos {{\alpha }^{T}}  \\
		\sin {{\beta }^{T}}\sin {{\alpha }^{T}}  \\
		\sin {{\beta }^{T}}  \\
	\end{matrix} \right]
\end{equation}

The Rx antenna distance vector can be expressed in the same way. In this system, there is one IRS and one AIRS, the IRS is in a fixed location, commonly placed on the surface of buildings, while the AIRS is in a moving position in the air, typically mounted on UAV or other aerial vehicles. The IRS and AIRS array consists of ${{N}_{\text{IRS}}}$ and ${{N}_{\text{AIRS}}}$ IRS units respectively. Since the frequency is determined at the same time, we assume that the IRS and AIRS have the same unit size of ${{\delta }_{h}}$ in the horizontal direction and ${{\delta }_{v}}$ in the vertical direction, respectively. The numbers of the arranged units in the horizontal and vertical directions of the AIRS are denoted by ${{H}_{A}}$ and ${{V}_{A}}$, respectively. For the IRS, they are denoted by ${{H}_{I}}$ and ${{V}_{I}}$, respectively. The angle of the IRS should also be taken into consideration, as concluded in \cite{IRSangle2021}, which proves the importance of the IRS angle for its performance in the system. As shown in Fig. 2 (b), IRS/AIRS have three degrees of freedom in different directions, we use ${\phi_{\text{IRS/AIRS}}}$, ${\theta_{\text{IRS/AIRS}}}$ and ${\psi_{\text{IRS/AIRS}}}$ to describe the pitch, yaw, and roll angles of IRS/AIRS, respectively (The roll angle is 0 for IRS). Then, taking IRS as example, the distance vector from the center of the IRS to the ($h, v$)-th ($h=1,2,3\ldots {{H}_{I}}$;$v=1,2,3\ldots {{V}_{I}}$) IRS unit can be expressed as
\begin{equation}
	\pmb{I}_{r}^{I}={{\left[ {{k}_{h}}{{\delta }_{h}},{{k}_{v}}{{\delta }_{v}},0 \right]}^{\text{T}}}\cdot \pmb{R}
\end{equation}
\begin{equation}
	\begin{aligned}
		 \pmb{R}= & \left( \begin{matrix}
			\cos \phi_{\text{IRS}}  & -\sin \phi_{\text{IRS}}  & 0  \\
			\sin \phi_{\text{IRS}}  & \cos \phi_{\text{IRS}}  & 0  \\
			0 & 0 & 1  \\
		\end{matrix} \right)\left( \begin{matrix}
			\cos \theta_{\text{IRS}}  & 0 & \sin \theta_{\text{IRS}}   \\
			0 & 1 & 0  \\
			-\sin \theta_{\text{IRS}}  & 0 & \cos \theta_{\text{IRS}}   \\
		\end{matrix} \right) \\ 
		& \left( \begin{matrix}
			1 & 0 & 0  \\
			0 & \cos \psi_{\text{IRS}}  & -\sin \psi_{\text{IRS}}   \\
			0 & \sin \psi_{\text{IRS}}  & \cos \psi_{\text{IRS}}   \\
		\end{matrix} \right) \\ 
	\end{aligned}
\end{equation}
where ${{k}_{h}}=\frac{2h-{{H}_{I}}-1}{2}$ and ${{k}_{v}}=\frac{2v-{{V}_{I}}-1}{2}$, the distance vectors from the center of the AIRS to other units can be expressed in the same way.

\begin{figure}[tbp]
	\centerline{\includegraphics[width=3.9in]{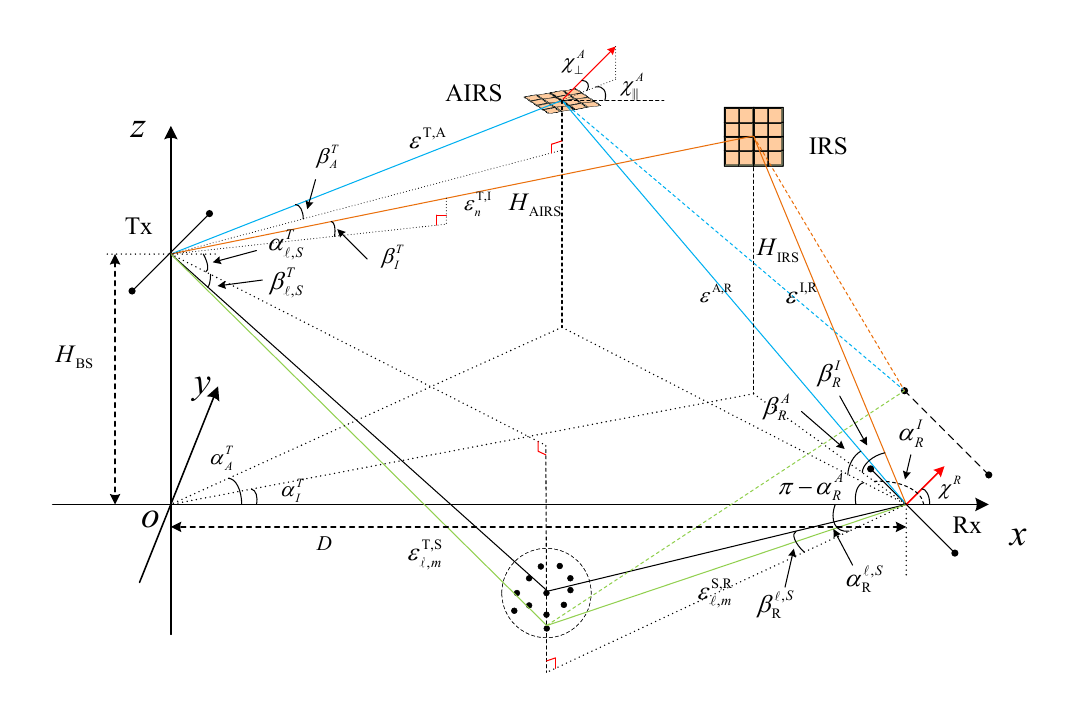}}
	\caption{A visualized illustration of the proposed 3D wideband channel model for IRS-assisted MIMO communications. For clarify, only one cluster are shown.}
	\label{AIRSmodel}
\end{figure}
\begin{figure}[tbp]\vspace{-3mm} 
	\centerline{\includegraphics[width=3in]{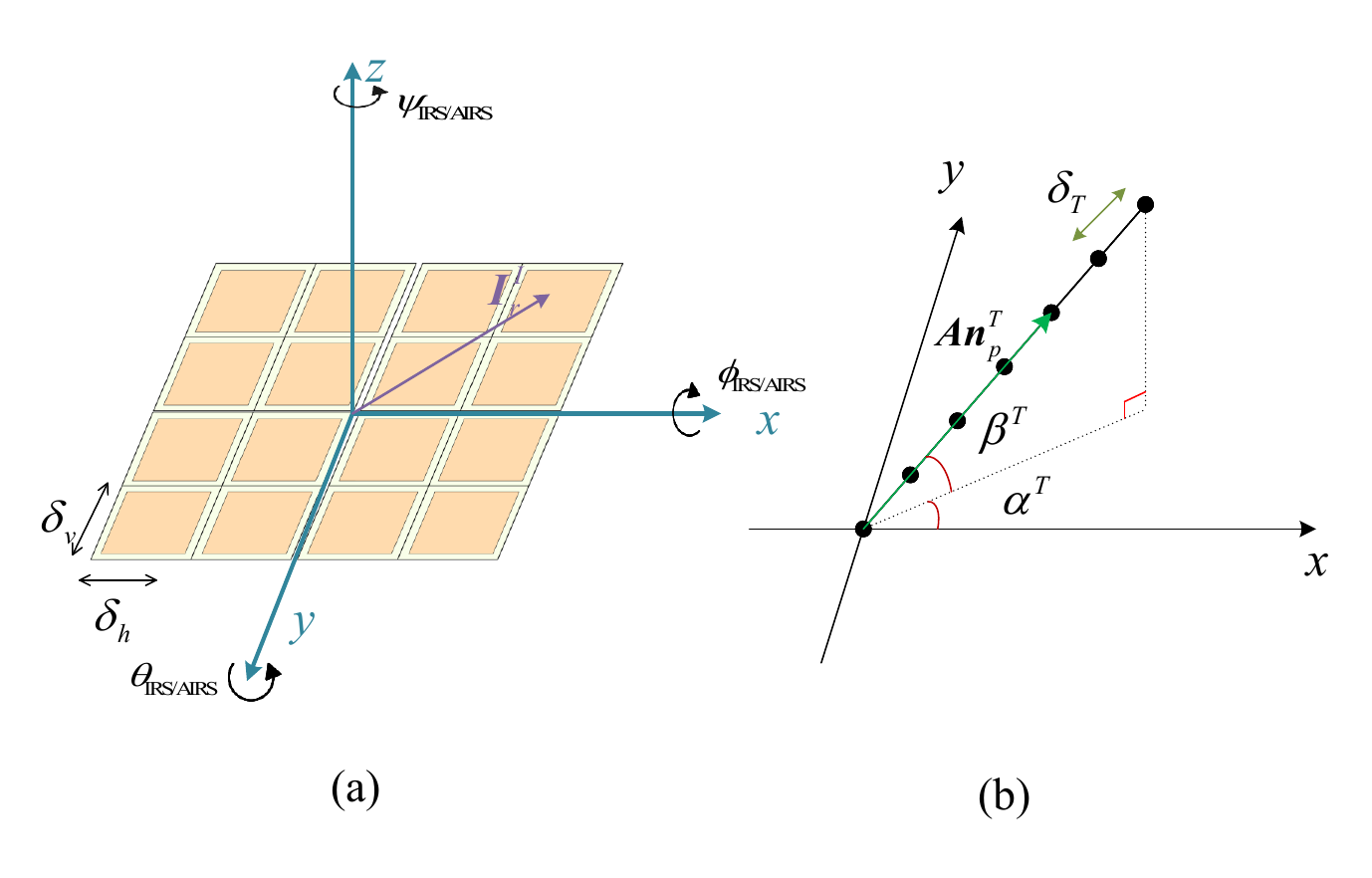}}
	\caption{Definitions of the 3D angles. (a) Pitch, yaw, and roll angles of the AIRS/IRS, (b) Azimuth and elevation angles of Tx/Rx ULAs.}\vspace{-5mm} 
	\label{angles}
\end{figure}

The UAV moves at the speed ${{v}_{\text{A}}}$ in the 3D space, thus the moving direction of AIRS should be described by the azimuth movement angle $\chi _{\parallel }^{A}$ and the elevation movement angle $\chi _{\bot }^{A}$. Whereas, the Rx only has an azimuth movement angle ${{\chi }^{R}}$ with speed ${{v}_{\text{R}}}$. The distance between Tx and Rx is $D$. ${{H}_{\text{BS}}}$, ${{H}_{\text{AIRS}}}$, and ${{H}_{\text{IRS}}}$ represent the heights of Tx, AIRS, and IRS, respectively. The abundant scattering will cause the multipath propagation of the signal, and increasing signal bandwidth will aggravate this effect. We use the CDL structure to simulate the frequency selectivity of the wideband channel, where each cluster is composed of multiple scatterers with similar distances and angles, forming a distinguishable path with a resolvable delay. In the proposed wideband channel model, we assume that there are $\mathcal{L}$ clusters between the Tx and Rx, in which every $\ell$-th ($\ell =1,2\ldots \mathcal{L}$) cluster have $M$ scatterers.

Based on the above geometry of the proposed model, three different propagation paths are considered: single-bounced at the AIRS side where waves scatter from the AIRS before arriving at Rx, single-bounced at the IRS side where waves scatter from the AIRS before arriving at Rx, and single-bounced at the Rx side (SBR) where waves scatter from the scatterers located around Rx before arriving at Rx. Hence, the CIR between the $p$-th Tx antenna and the $q$-th Rx antenna in MIMO channels can be expressed as follows:
\begin{equation}
  {{h}_{pq}}(t,\tau )=h_{pq}^{\text{AIRS}}(t,\tau )+h_{pq}^{\text{IRS}}(t,\tau )+h_{pq}^{\text{SBR}}(t,\tau )
\end{equation}
where $t$ indicates time. We further have
\begin{equation}\label{hAIRS}
	\begin{aligned}
	h_{pq}^{\text{AIRS}}(t,\tau )&=\sqrt{\frac{{{K}_{\text{AIRS}}}}{{{K}_{\text{Rice}}}+1}}\frac{1}{\sqrt{{{N}_{\text{AIRS}}}}}\sum\limits_{{{n}_{a}}=1}^{{{N}_{\text{AIRS}}}}{{{e}^{j{{\varphi }_{{{n}_{a}}}}(t)}}}\\
	& \times {{e}^{-j\frac{2\pi }{\lambda }(\varepsilon _{{{n}_{a}}}^{\text{T,A}}(t)+\varepsilon _{{{n}_{a}}}^{\text{A,R}}(t))}} \\ 
	& \times {{e}^{j\frac{2\pi }{\lambda }{{v}_{\text{A}}}t(\cos \beta _{\text{A}}^{\text{T}}(t)\cos \chi _{\bot }^{\text{A}}\cos (\alpha _{\text{A}}^{\text{T}}(t)-\chi _{\parallel }^{\text{A}})+\sin \beta _{\text{A}}^{\text{T}}(t)\sin \chi _{\bot }^{\text{A}})}} \\ 
	& \times {{e}^{-j\frac{2\pi }{\lambda }{{v}_{\text{A}}}t(\cos \beta _{\text{R}}^{\text{A}}(t)\cos \chi _{\bot }^{\text{A}}\cos (\alpha _{\text{R}}^{\text{A}}(t)-\chi _{\parallel }^{\text{A}})+\sin \beta _{R}^{A}(t)\sin \chi _{\bot }^{\text{A}})}} \\ 
	& \times {{e}^{j\frac{2\pi }{\lambda }{{v}_{R}}t\cos (\alpha _{\text{R}}^{\text{A}}(t)-{{\chi }^{R}})\cos \beta _{\text{R}}^{\text{A}}(t)}}\delta (\tau -{{\tau }^{\text{AIRS}}}(t)) \\ 
   \end{aligned}
\end{equation}
\begin{equation}\label{hIRS}
	\begin{aligned}
	h_{pq}^{\text{IRS}}(t,\tau )&=\sqrt{\frac{{{K}_{\text{IRS}}}}{{{K}_{\text{Rice}}}+1}}\frac{1}{\sqrt{{{N}_{\text{IRS}}}}}\sum\limits_{{{n}_{i}}=1}^{{{N}_{\text{IRS}}}}{{{e}^{j{{\varphi }_{{{n}_{i}}}}(t)}}}\\
	& \times {{e}^{-j\frac{2\pi }{\lambda }(\varepsilon _{{{n}_{i}}}^{\text{T,I}}+\varepsilon _{{{n}_{i}}}^{\text{I,R}}(t))}}\\
	& \times {{e}^{j\frac{2\pi }{\lambda }{{v}_{R}}t\cos (\alpha _{\text{R}}^{\text{I}}(t)-{{\chi }^{R}})\cos \beta _{\text{R}}^{\text{I}}(t)}}\delta (\tau -{{\tau }^{\text{IRS}}}(t))
    \end{aligned}
\end{equation}
where ${{\varphi }_{{{n}_{a}}}}(t)$ and ${{\varphi }_{{{n}_{i}}}}(t)$ denote the phase shifts at the $n$-th AIRS and $n$-th IRS unit, respectively. $\lambda $ is the wavelength. ${{K}_{\text{Rice}}}$ denotes the Ricean factor, weight factor ${{K}_{\text{AIRS}}}$ and ${{K}_{\text{IRS}}}$ satisfy ${{K}_{\text{AIRS}}}+{{K}_{\text{IRS}}}={{K}_{\text{Rice}}}$. In addition, $\varepsilon _{{{n}_{a}}}^{\text{T,A}}(t)$, $\varepsilon _{{{n}_{a}}}^{\text{A,R}}(t)$, $\varepsilon _{{{n}_{i}}}^{\text{T,I}}$, $\varepsilon _{{{n}_{i}}}^{\text{I,R}}(t)$ account for the distances of $An_{(p)}^{T}\!-\!\text{AIR}{{\text{S}}^{(n)}}$, $\text{AIR}{{\text{S}}^{(n)}}\!-\!An_{(q)}^{R}$, $An_{(p)}^{T}\!-\!\text{IR}{{\text{S}}^{(n)}}$, $\text{IR}{{\text{S}}^{(n)}}\!-\!An_{(q)}^{R}$, respectively, where $\text{AIRS/IR}{{\text{S}}^{(n)}}$ represent the $n$-th unit of AIRS/IRS. In this paper, bold variables represent vectors, and the distance term can be obtained by taking the Euclidean norm of the distance vector. The distance terms in (\ref{hAIRS}) and (\ref{hIRS}) can be derived as $\pmb{\varepsilon} _{{{n}_{a}}}^{\text{A,R}}={\pmb{\varepsilon }^{\text{A,R}}}(t)+\pmb{An}_{q}^{R}-\pmb{I}_{r}^{\text{A}}$, $\pmb{\varepsilon} _{{n}_{i}}^{\text{I,R}}(t)={\pmb{\varepsilon }^{\text{I,R}}}(t)+\pmb{An}_{q}^{R}-\pmb{I}_{r}^{I}$, by substituting $\pmb{An}_{q}^{R}$ with $\pmb{An}_{p}^{T}$, ${\pmb{\varepsilon }^{\text{A,R}}}$ and ${\pmb{\varepsilon }^{\text{I,R}}}$ with ${\pmb{\varepsilon }^{\text{T,A}}}$ and ${\pmb{\varepsilon }^{\text{T,I}}}$, we can obtain $\pmb{\varepsilon} _{{{n}_{a}}}^{\text{T,A}}$ and $\pmb{\varepsilon} _{{n}_{i}}^{\text{T,I}}$.
\begin{equation}
	{\pmb{\varepsilon }^{\text{T,I}}}={{[{{x}_{\text{I}}},{{y}_{\text{I}}},{{H}_{\text{IRS}}}-{{H}_{\text{BS}}}]}^{\text{T}}}
\end{equation}
\begin{equation}
	{\pmb{\varepsilon }^{\text{I,R}}}(t)={{[D\!+\!{{v}_{R}}t\cos {{\chi }^{R}}\!-\!{{x}_{\text{I}}},{{v}_{R}}t\sin {{\chi }^{R}}\!-\!{{y}_{\text{I}}},-{{H}_{\text{IRS}}}]}^{\text{T}}}
\end{equation}
\begin{equation}
	\left\{ \begin{matrix}
		{{x}_{\text{A}}}(t)={{x}_{\text{A}}}(0)+{{v}_{\text{A}}}t\sin \chi _{\bot }^{\text{A}}\cos \chi _{\parallel }^{\text{A}}  \\
		{{y}_{\text{A}}}(t)={{y}_{\text{A}}}(0)+{{v}_{\text{A}}}t\sin \chi _{\bot }^{\text{A}}\sin \chi _{\parallel }^{\text{A}}  \\
		{{z}_{\text{A}}}(t)={{H}_{\text{AIRS}}}+{{v}_{\text{A}}}t\cos \chi _{\bot }^{\text{A}}-{{H}_{\text{BS}}}  \\
	\end{matrix} \right.
\end{equation}
\begin{equation}
	{\pmb{\varepsilon }^{\text{T,A}}}(t)={{[{{x}_{\text{A}}}(t),{{y}_{\text{A}}}(t),{{z}_{\text{A}}}(t)]}^{\text{T}}}
\end{equation}
\begin{equation}
	{\pmb{\varepsilon }^{\text{A,R}}}(t)=\left[ \begin{matrix}
		D+{{v}_{R}}t\cos {{\chi }^{R}}-{{x}_{\text{A}}}(t)  \\
		{{v}_{R}}t\sin {{\chi }^{R}}-{{y}_{\text{A}}}(t)  \\
		-({{H}_{\text{AIRS}}}+{{v}_{\text{A}}}t\cos \chi _{\bot }^{\text{A}})  \\
	\end{matrix} \right]
\end{equation}
where ${{x}_{\text{A}}}(0)$, ${{y}_{\text{A}}}(0)$, and ${{z}_{\text{A}}}(0)$ represent the coordinates of AIRS at the initial time, while ${{x}_{\text{A}}}(t)$, ${{y}_{\text{A}}}(t)$, and ${{z}_{\text{A}}}(t)$ represent the coordinates of AIRS at time $t$. the AAoD, AAoA, EAoD, and EAoA of the AIRS path for Tx-Rx link are given by
\begin{equation}\vspace{-1mm} 
	\alpha _{\text{A}}^{\text{T}}(t)=\arctan \left( {{x}_{\text{A}}}(t)/{{y}_{\text{A}}}(t) \right)
\end{equation}
\begin{equation}\vspace{-1mm} 
	\beta _{\text{A}}^{\text{T}}(t)=\arcsin \left( {{z}_{\text{A}}}(t)/\left\| {{\varepsilon }^{\text{T,A}}}(t) \right\| \right)
\end{equation}
\begin{equation}\vspace{-1mm} 
	\beta _{\text{R}}^{\text{A}}(t)=\arcsin \frac{{{H}_{\text{AIRS}}}+{{v}_{\text{A}}}t\cos \chi _{\bot }^{\text{A}}}{\left\| {\pmb{\varepsilon }^{\text{A,R}}}(t) \right\|}
\end{equation}
\begin{equation}\vspace{-1mm} 
	\alpha _{\text{R}}^{\text{A}}(t)=\arcsin \frac{{{y}_{\text{A}}}(t)-{{v}_{R}}t\sin {{\chi }^{R}}}{\left\| {\pmb{\varepsilon }^{\text{A,R}}}(t) \right\|\cos \beta _{\text{R}}^{\text{A}}(t)}
\end{equation}

Similarly, the EAoA and AAoA of the IRS path for IRS-Rx link can be derived as
\begin{equation}
	\beta _{\text{R}}^{\text{I}}(t)=\arcsin \left( {{\text{H}}_{\text{IRS}}}/\left\| {\pmb{\varepsilon }^{\text{I,R}}}(t) \right\| \right)
\end{equation}
\begin{equation}
	\alpha _{\text{R}}^{\text{I}}(t)=\arcsin \frac{{{y}_{\text{I}}}-{{v}_{R}}t\sin {{\chi }^{R}}}{\left\| {\pmb{\varepsilon }^{\text{I,R}}}(t) \right\|\cos \beta _{\text{R}}^{\text{I}}(t)}
\end{equation}

Then, the SBR component $h_{pq}^{\text{SBR}}(t,\tau )$ can be expressed as
\begin{equation}\label{hSBR}
	\begin{aligned}
	h_{pq}^{\text{SBR}}(t,\tau )&=\sqrt{\frac{1}{{{K}_{\text{Rice}}}+1}}\sqrt{\frac{{{P}_{(\ell )}}(t)}{\mathcal{L}}}\frac{1}{\sqrt{M}}\sum\limits_{\ell =1}^{\mathcal{L}}{\sum\limits_{m=1}^{M}{{{e}^{j{{\varphi }_{l,m}}}}}}\\
	& \times {{e}^{-j\frac{2\pi }{\lambda }(\varepsilon _{\ell ,m}^{\text{T,S}}+\varepsilon _{\ell ,m}^{\text{S,R}}(t))}}\\
	& \times {{e}^{j\frac{2\pi }{\lambda }{{v}_{R}}t\cos (\alpha _{\text{R}}^{\ell ,\text{S}}(t)-{{\chi }^{R}})\cos \beta _{\text{R}}^{\ell ,\text{S}}(t)}}\delta (\tau -{{\tau }^{{{\ell }_{1}}}}(t))
 	\end{aligned}
\end{equation}
where ${{\varphi }_{l,m}}\in \left[ 0,2\pi  \right)$ denotes the independent distributed random phases, and $\mathcal{L}$ is the total  number of clusters, and ${{P}_{(\ell )}}(t)$ denotes the cluster power, which can be calculated as \cite{3gpp.38.901}
\begin{equation}
	{{P}_{\ell }}(t)=\exp (-{{\tau }_{\ell }}(t)\frac{{{r}_{\tau }}-1}{{{r}_{\tau }}{{\sigma }_{\tau }}})\cdot {{10}^{\frac{-{{Z}_{\ell }}}{10}}}
\end{equation}
where ${{\tau }_{\ell }}$ is the delay of the $\ell$-th cluster, ${{r}_{\tau }}$ is the delay scaling function related to the scene, ${{\sigma }_{\tau }}$ is the delay spread, ${{Z}_{\ell }}$ is a random variable following a Gaussian distribution satisfy ${{Z}_{\ell }}\sim N(0,{{\zeta }^{2}})$. Through ${{P}_{\ell }}=\frac{{{{{P}'}}_{\ell }}}{\sum\nolimits_{\ell =1}^{\mathcal{L}}{{{{{P}'}}_{\ell }}}}$, ${{P}_{(\ell )}}(t)$ is normalized to satisfy $\sum\nolimits_{\ell =1}^{\mathcal{L}}{{{P}_{\ell }}(t)}=1$. The distance terms in (\ref{hSBR}) can be derived as
\begin{equation}
\begin{aligned}
	\varepsilon _{\ell ,m}^{\text{S,R}}(t)&=({{(\varepsilon _{\ell ,m}^{\text{S,R}}(0)\cos \beta _{\text{R}}^{\ell ,\text{S}}(0)\cos \alpha _{\text{R}}^{\ell ,\text{S}}(0)+{{v}_{R}}t\cos {{\chi }^{R}})}^{2}} \\ 
	& +{{(\varepsilon _{\ell ,m}^{\text{S,R}}(0)\cos \beta _{\text{R}}^{\ell ,\text{S}}(0)\sin \alpha _{\text{R}}^{\ell ,\text{S}}(0)+{{v}_{R}}t\sin {{\chi }^{R}})}^{2}} \\ 
	& +{{(\varepsilon _{\ell ,m}^{\text{S,R}}(0)\sin \beta _{\text{R}}^{\ell ,\text{S}}(0))}^{2}}{{)}^{1/2}} \\ 
\end{aligned}
\end{equation}
\begin{equation}\vspace{-1mm} 
\begin{aligned}
	\varepsilon _{\ell ,m}^{\text{T,S}}(t)&=({{(\varepsilon _{\ell ,m}^{\text{S,R}}(t)\cos \beta _{\text{R}}^{\ell ,\text{S}}(t)\sin \alpha _{\text{R}}^{\ell ,\text{S}}(t))}^{2}} \\ 
	& +{{(D-\varepsilon _{\ell ,m}^{\text{S,R}}(t)\cos \beta _{\text{R}}^{\ell ,\text{S}}(t)\cos \alpha _{\text{R}}^{\ell ,\text{S}}(t))}^{2}} \\ 
	& +{{({{H}_{\text{BS}}}-\varepsilon _{\ell ,m}^{\text{S,R}}(t)\sin \beta _{\text{R}}^{\ell ,\text{S}}(t))}^{2}}{{)}^{1/2}} \\ 
\end{aligned}
\end{equation}

Moreover, it can be seen from Fig. 1 that the EAoA and AAoA of the SBR path have geometric relationships of the $\varepsilon _{\ell ,m}^{\text{S,R}}(t)$ and $\varepsilon _{\ell ,m}^{\text{T,S}}(t)$.
\begin{equation}
	\beta _{\text{R}}^{\ell ,\text{S}}(t)=\arcsin (\varepsilon _{\ell ,m}^{\text{S,R}}(0)\sin \beta _{\text{R}}^{\ell ,\text{S}}(0)/\varepsilon _{\ell ,m}^{\text{S,R}}(t))
\end{equation}
\begin{equation}
	\alpha _{\text{R}}^{\ell ,\text{S}}(t)=\arcsin \frac{\varepsilon _{\ell ,m}^{\text{S,R}}(0)\cos \beta _{\text{R}}^{\ell ,\text{S}}(0)\sin \alpha _{\text{R}}^{\ell ,\text{S}}(0)}{\varepsilon _{\ell ,m}^{\text{S,R}}(t)\cos \beta _{\text{R}}^{\ell ,\text{S}}(t)}
\end{equation}
\subsection{The Distribution of Scatterers}
Scatterers are distributed in the form of clusters anisotropically. In the cluster structure, especially for mmWave communications, the directions of the rays in each cluster are generally limited to a certain range\cite{xiong20223d}. Here we adopt the truncated Gaussian probability density function (PDF) to characterize the distribution of arrival angles $\alpha _{\text{R}}^{\ell ,\text{S}}$ and $\beta _{\text{R}}^{\ell ,\text{S}}$ \cite{wu2017general,xiong2021statistical},
\begin{equation}
	f(\theta ,{{\mu }_{\theta }},{{\sigma }_{\theta }},{{\theta }_{\text{low}}},{{\theta }_{\text{up}}})=\frac{\phi ({{\mu }_{\theta }},\sigma _{\theta }^{2};\theta )}{\Phi ({{\mu }_{\theta }},\sigma _{\theta }^{2};{{\theta }_{\text{up}}})-\Phi ({{\mu }_{\theta }},\sigma _{\theta }^{2};{{\theta }_{\text{low}}})}
\end{equation}
where ${{\mu }_{\theta }}$ and ${{\sigma }_{\theta }}$ denote the mean value and variance of the signal direction $\theta $, respectively; ${{\theta }_{\text{low}}}$ and ${{\theta }_{\text{up}}}$ are the lower and upper bounds of the truncated Gaussian distributed signal direction $\theta $. Moreover, $\phi (\cdot )$ and $\Phi (\cdot )$ are the PDF and cumulative distribution functions (CDF) of the standard Gaussian distribution.
\subsection{IRS phase shifts design}
In this section, the objective is to jointly design the phases of AIRS and IRS to improve the statistical features of the channel. Initially, two simple methods are introduced as benchmarks. Subsequently, a new approach is proposed to attain an equivalent LoS path by dynamically adjusting the phases of the electromagnetic waves that reach the IRS units. Lastly, the case of non-ideal IRS is discussed.
\begin{itemize}
\item \emph{Method 1}: In this method, we set ${{\varphi }_{{{n}_{i}}}}(t)=0$ for every unit of IRS and ${{\varphi }_{{{n}_{a}}}}(t)=0$ for AIRS, i.e, without IRS.
\item \emph{Method 2}: In this method, IRS units are considered as general scatterers, the IRS controller adopts a constant phase configuration, which ${{\varphi }_{{{n}_{i}}}}(t)$ and ${{\varphi }_{{{n}_{a}}}}(t)$ are independent and randomly distributed values in $\left[ 0,2\pi  \right)$.
\item \emph{Method 3}: In this method, we aim to align the waves from two IRSs at the Rx to act as a virtual LoS path. Firstly, we determine the phase of each IRS unit as $\varphi _{hv}^{\text{IRS}}-{{k}_{0}}(\varepsilon _{{{n}_{i}}}^{\text{T,I}}+\varepsilon _{{{n}_{i}}}^{\text{I,R}}(t))=\phi $, where ${{k}_{0}}=\frac{2\pi }{\lambda }$ is the wave number, and $\phi$ is a constant value representing the desired phase at the receiver end. Furthermore, we determine the phase of each unit in AIRS as $\varphi _{hv}^{\text{AIRS}}={{k}_{0}}(\varepsilon_{{{n}_{a}}}^{\text{T,A}}(t)+\varepsilon _{{{n}_{a}}}^{\text{A,R}}(t))+\phi $. Note that the complex-exponential function ${{e}^{j\phi }}$ is periodic with $2\pi$, thus the actual phase shifts provided by IRS units should be mod $2\pi$.
\item \emph{Method 4}: In this method, non-idealized IRS will be considered. The phase of IRS in the real world is discontinuous, and Method 4 is a discretization process based on Method 3. The process of discretization is to divide $[0,2\pi )$ into ${{2}^{n}}$ intervals, where $n$ is the number of quantized bits. The value of ${{\varphi }_{{{n}_{i}}}}(t)$ and ${{\varphi }_{{{n}_{a}}}}(t)$ is the median value of the nearest equal interval.
\end{itemize}
\section{Statistical Characteristics of the Channel Models}
\subsection{Space-Time Correlation Function (STCF)}
The spatial-temporal correlation function of a channel can be used to evaluate the correlation of the channel as it varies over time and space. The normalized spatial-temporal cross-correlation functions of the proposed channel model can be derived from two different  complex channel coefficients ${{h}_{pq}}(t)$ and ${{h}_{{p}'{q}'}}(t)$.
\begin{equation}\label{ACFCCF}
	\begin{aligned}
	{{\rho }_{(pq),\!({p}'\!{q}')}}(t;\!\Delta p,\!\Delta q,\!\Delta t)\!=\!\frac{\mathbb{E}[h_{pq}^{*}(t){{h}_{{p}'\!{q}'}}(t+\Delta t)]}{\sqrt{\mathbb{E}\!\left[\!{{\left\| {{h}_{pq}}(t)\!\right\|}^{2}} \!\right]\!\mathbb{E}\left[\!{{\left\| {{h}_{{p}'\!{q}'}}(t\!+\!\Delta t)\!\right\|}^{2}}\!\right]}}
	\end{aligned}
\end{equation}
where $\Delta t$ denotes the time difference, $\mathbb{E}[\cdot ]$ is the statistical average, the superscript ${{[\cdot ]}^{*}}$ is the complex conjugation operator. $\Delta p=\left| {p}'-p \right|{{\delta }_{T}}/\lambda $ is the normalized antenna spacing between the $p$-th and ${p}'$-th Tx antenna, and $\Delta q=\left| {q}'-q \right|{{\delta }_{R}}/\lambda $ is the normalized antenna spacing between the $q$-th and ${q}'$-th Rx antenna. The temporal auto-correlation function (ACF) of the proposed channel model can be obtained by imposing $p={p}'$ and $q={q}'$ in (\ref{ACFCCF}), then, by imposing $\Delta t=0$, we can get the spatial cross-correlation function (CCF).
\subsection{Frequency Cross-correlation Function (FCF)}
By performing Fourier transform on CIR, we can obtain the frequency response ${H}_{pq}(t,f)$ of the channel, the FCF of the proposed model is given by
\begin{equation}
	{{\rho }_{pq}}(t,\Delta f)=\frac{\mathbb{E}[H_{pq}^{*}(t,f){{H}_{pq}}(t,f+\Delta f)]}{\sqrt{\mathbb{E}\left[ {{\left\| {{H}_{pq}}(t,f) \right\|}^{2}} \right]\mathbb{E}\left[ {{\left\| {{H}_{pq}}(t,f\!+\!\Delta f) \right\|}^{2}} \right]}}
\end{equation}

\subsection{Channel Capacity}
The instantaneous MIMO channel capacity of the proposed model, under the assumption that given perfect channel information at the Rx, can be expressed as \cite{Patzold2005}
\begin{equation}
	C(t)={{\log }_{2}}\left( \det \left( {{\mathbf{I}}_{P}}+\frac{\rho }{Q}\mathbf{H}_{pq}(t){{\mathbf{H}}^{H}_{pq}}(t) \right) \right)
\end{equation}
where$P>Q$,   denotes the matrix determinant, ${{\mathbf{I}}_{P}}$ is the $P\times P$ identity matrix, $\rho $ is the average signal-to-noise ratio (SNR), ${{(\cdot )}^{\text{H}}}$ denotes the transpose conjugate operation. Since the time-varying channel matrix ${\mathbf{H}_{pq}}(t)=\left[ {{h}_{pq}}(t) \right]$ is deterministic of time $t$, the time average capacity of the MIMO channel can be expressed as
\begin{equation}
	{{C}_\text{mean}}=\underset{T\to \infty }{\mathop{\lim }}\,\frac{1}{2T}\int_{-T}^{T}{C(t)}dt
\end{equation}

\section{Numerical Analysis}
In this section, the impacts of the proposed methods on correlation function, and channel capacity are analyzed via simulations. The basic model parameters are listed here or specified otherwise:${{f}_{c}}=2.4\text{GHz}$, $L=16$, $M=20$, ${{\delta }_{v}}={{\delta }_{h}}=0.5\lambda $, ${{K}_{\text{AIRS}}}={{K}_{\text{IRS}}}=0$dB, ${{\alpha }^{T}}=\frac{\pi }{5}$, ${{\beta }^{T}}=\frac{\pi }{3}$, ${{\alpha }^{R}}=\frac{\pi }{3}$, ${{\beta }^{R}}=\frac{\pi }{12}$. For clusters, $\varepsilon _{\ell ,m}^{\text{S,R}}(0)=60$m, $\alpha _{\text{R}}^{\ell ,\text{S}}$ and $\beta _{\text{R}}^{\ell ,\text{S}}$ follow the same distribution as mentioned in last section where ${{\mu }_{\theta }}=\frac{\pi }{6}$, ${{\sigma }_{\theta }}=\frac{\pi }{18}$, ${{\theta }_{\text{low}}}=\frac{\pi }{12}$, ${\theta_{\text{IRS}}}=\frac{\pi }{2}$ and ${{\theta }_{\text{up}}}=\frac{\pi }{3}$. The IRS is located at $({{x}_{\text{I}}},{{y}_{\text{I}}},{{H}_{\text{IRS}}})=(\text{100m, 50m, 10m})$ with the rotation angles ${{\psi }_{\text{IRS}}}=\frac{\pi }{6}$, ${\theta_{\text{IRS}}}=\frac{\pi }{2}$ and ${{\phi }_{\text{IRS}}}=\frac{\pi }{2}$, and the AIRS is located at $({{x}_{\text{A}}}, {{y}_{\text{A}}}, {{H}_{\text{AIRS}}})=(\text{75m, 75m, 30m})$ with the rotation angles ${{\psi }_{\text{AIRS}}}=\frac{\pi }{2}$, ${\theta_{\text{AIRS}}}=\frac{\pi }{2}$ and ${{\phi }_{\text{AIRS}}}=0$. Two IRSs have the same unit configuration, the units of two IRSs are set in the same size as ${{\delta }_{h}}={{\delta }_{v}}=\frac{\lambda }{2}$ and ${{H}_{I}}={{V}_{I}}={{H}_{A}}={{V}_{A}}=10$. The moving speeds and directions of the AIRS and Rx are set as ${{v}_{\text{A}}}=5\text{m/s}$, $\chi _{\parallel }^{A}=\chi _{\bot }^{A}=\frac{\pi }{10}$, ${{v}_{\text{R}}}=30\text{m/s}$, ${{\chi }^{R}}=\frac{\pi }{6}$, respectively.

\begin{figure}[bp]\vspace{-5mm} 
	\centerline{\includegraphics[width=3.5in]{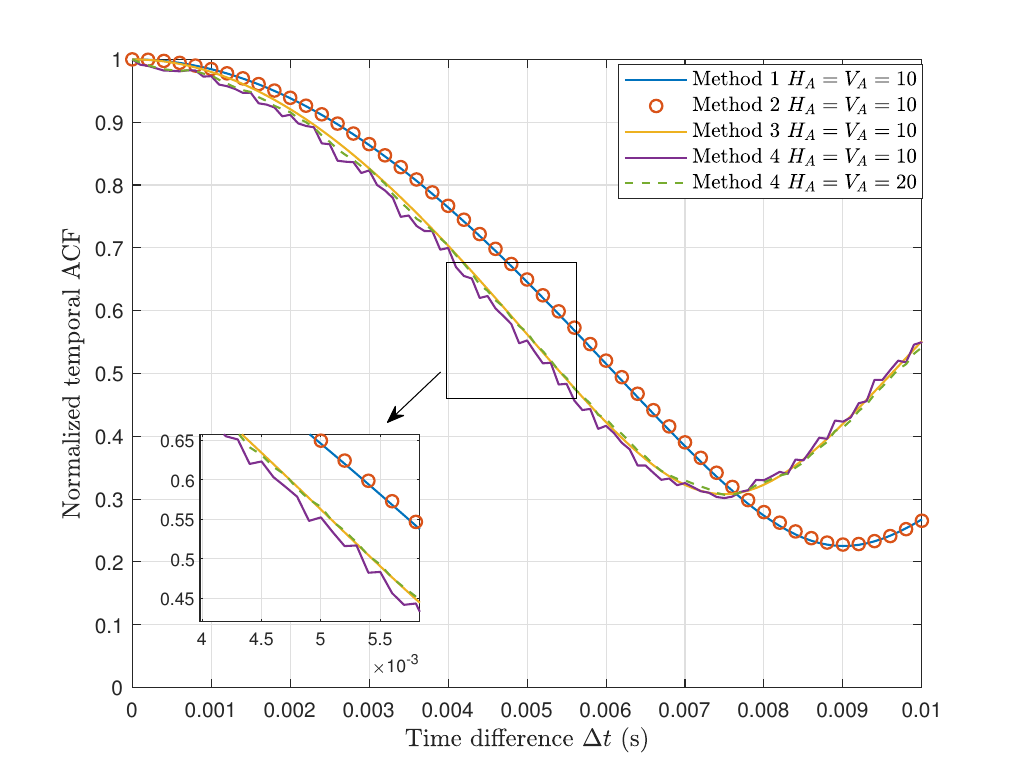}}
	\caption{Temporal ACF of the proposed channel model with different phase shifts methods at $t=0$s (Method 4 in 4-bits, ${{H}_{I}}={{V}_{I}}={{H}_{A}}={{V}_{A}}$).}
	\label{fig3}
\end{figure}

Fig. 3 demonstrates the influence of both the proposed phase shifts methods and the quantity of IRS units on the temporal ACF. Method 1 and Method 2 yield comparable results, while Method 3 exhibits a similar outcome to that of Method 4. Due to the discrete phase shifts nature of Method 4, the ACF exhibits noticeable fluctuations. By increasing the number of IRS units, the correlation time of the channel in Method 4 exhibits a greater proximity to that of Method 3, meaning that the channel estimation frequency will decrease.

Fig. 4 presents the impact of the proposed phase shifts method and the size of the IRS elements on the spatial CCF. Similar to ACF curves shown in Fig. 3, the outcomes of Method 1 and Method 2 are comparable while Method 1 yields similar results to Method 4. Method 3 achieves a similar LoS effect by aligning the phases of the two IRSs, and the value of the CCF stabilizes at $\frac{{{K}_{\text{Rice}}}}{{{K}_{\text{Rice}}}+1}$ after a certain point. Meanwhile, the influence of the size of the IRS units on the temporal ACF is negligible.

Fig. 5 illustrates the impact of the proposed phase shifts method on the spatial CCF with respect to different times. Simulation results show the non-stationarity of the channel in the time domain. Regarding Method 1, the influence of time domain non-stationarity is notable evidenced by a fluctuating trend observed in the CCF. On the other hand, while time-domain non-stationarity also affects Method 3, it initially experiences fluctuations but eventually plateaus at a constant value of $\frac{{{K}_{\text{Rice}}}}{{{K}_{\text{Rice}}}+1}$.

\begin{figure}[tbp]\vspace{-5mm} 
	\centerline{\includegraphics[width=3.5in]{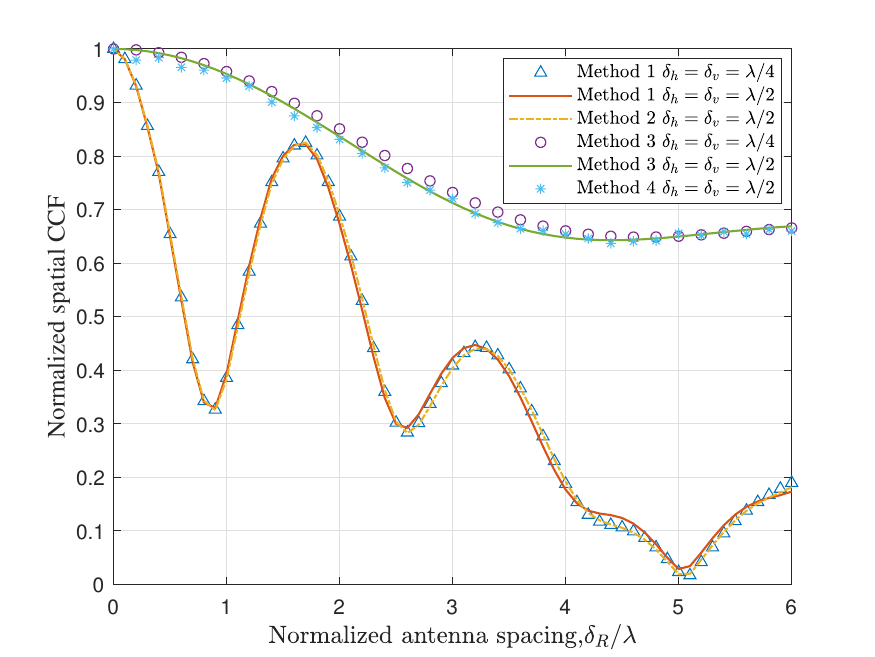}}
	\caption{Spatial CCF of the proposed channel model under different IRS unit sizes (Method 4 in 3-bits, $t=0.5$s).}
	\label{fig4}
\end{figure}
\begin{figure}[tbp]\vspace{-5mm} 
	\centerline{\includegraphics[width=3.5in]{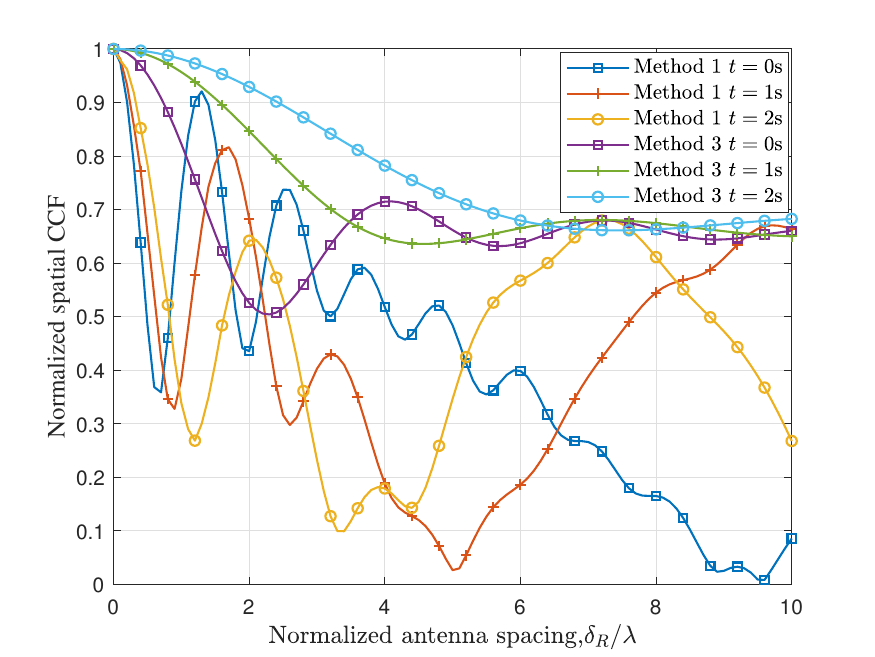}}
	\caption{Spatial CCF of the proposed channel model with two typical phase shifts methods under different time instants.}\vspace{-3mm} 
	\label{fig5}
\end{figure}
\begin{figure}[tbp]
	\centerline{\includegraphics[width=3.5in]{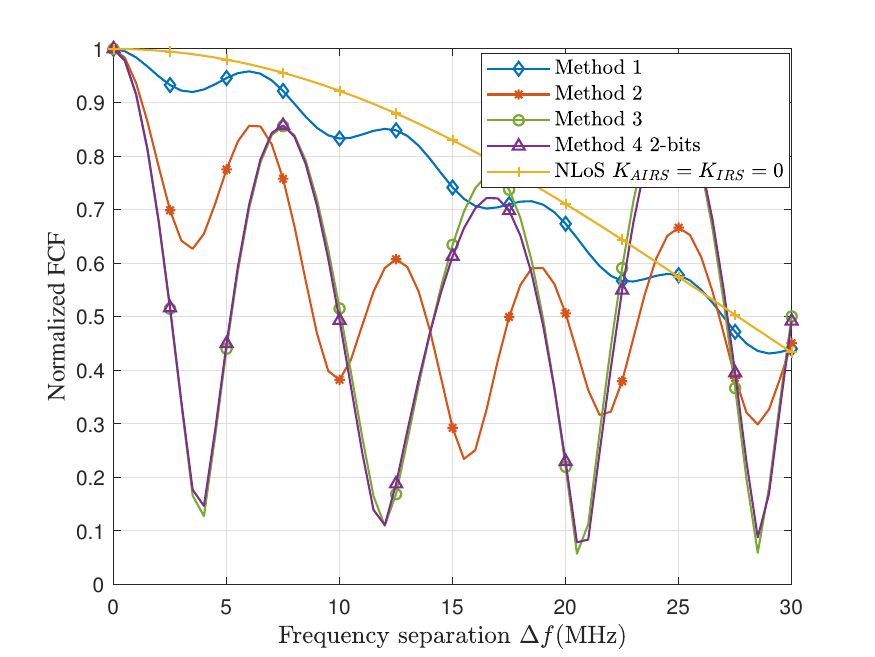}}
	\caption{FCF of the proposed channel model with different phase shifts methods at $t=0.5$s.}\vspace{-3mm} 
	\label{fig6}
\end{figure}

Fig. 6 portrays the impact of the proposed phase shifts method on the frequency-domain FCF. The results indicate that, with Method 1, the channel FCF experiences a slow decay similar to that of the NLoS path, leading to an increase in the coherence bandwidth. Method 2 performs better than Method 1. Meanwhile, the 2-bits quantization utilized in Method 4 can effectively approximate the results achieved by Method 3. Both Method 3 and Method 4 can achieve low frequency correlation. The frequency characteristics of the channel have low sensitivity to non-ideal IRS, which is in line with the conclusion in \cite{xiong20223d}.

\begin{figure}[tbp]\vspace{-3mm} 
	\centerline{\includegraphics[width=3.5in]{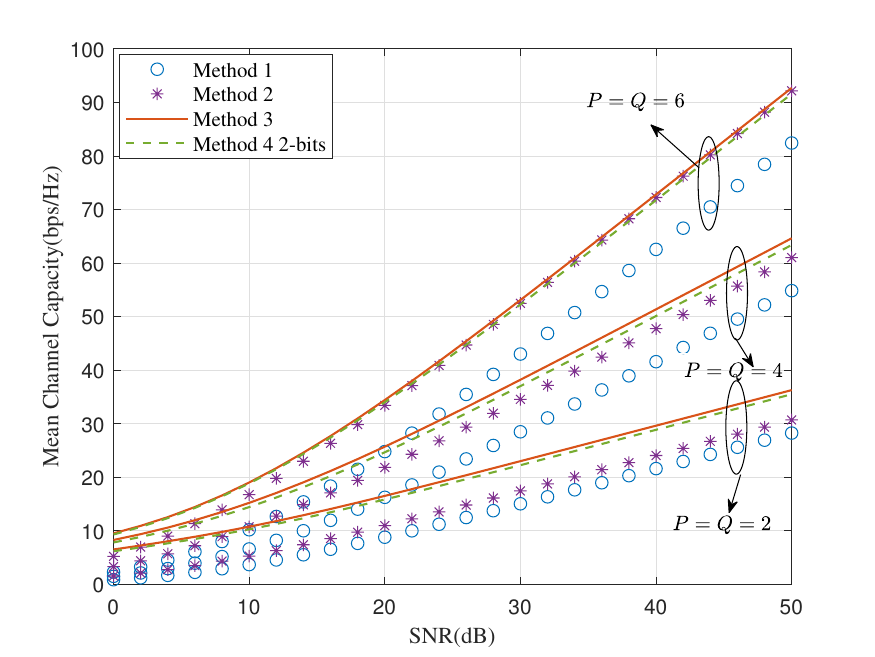}}
	\caption{Impact of the proposed methods on the channel capacity under the different numbers of antennas.}\vspace{-3mm} 
	\label{fig7}
\end{figure}

Fig. 7 shows the mean channel capacity of the various methods against SNR for different numbers of antennas at both the Tx and Rx. Under low SNR conditions, Method 3 and Method 4 have significant advantages over the other two methods. Under high SNR conditions, the remaining three methods outperform Method 1 significantly, as the number of antennas decreases, the disparity between Method 2 and Method 3\&4 in terms of channel capacity widens gradually. It can be seen from Method 3 and Method 4 that the non-idealized IRS has little impact on the channel capacity of the MIMO system.

\section{Conclusion}
A novel 3D wideband simulation channel model has been proposed for MIMO communication systems assisted by AIRS and IRS. The scatterers in the environment are modeled as clusters, and the mobility of AIRS is considered. Based on this model, the CIR, correlation function, and channel capacity are derived and analyzed. Furthermore, a novel method for designing phase shiftss of multiple IRSs is developed. Simulation results show that the deployment of AIRS and IRS can effectively improve the statistical characteristics of the channel, leading to an increase in channel capacity. Additionally, non-ideal IRS can also achieve good results  demonstrating the promising prospects of the IRS in practical applications. These results provide valuable references for the design of AIRS-assisted MIMO systems.
\section*{Acknowledgment}
This work was supported by The Open Research Project Programme of the State Key Laboratory of Internet of Things for Smart City (University of Macau): SKL-IoTSC(UM)-2021-2023/ORPF/SA02/2022. And the authors would like to thank Sungrow Power Supply Co., Ltd. for the support of application senairos.
\bibliographystyle{IEEEtran}  
\bibliography{bibtex}  

\end{document}